\documentstyle[psfig,aaspp4]{article}

\begin{document}

\title{Discovery of Radio-loud Broad Absorption Line Quasars \\ 
Using Ultraviolet Excess and Deep Radio Selection\altaffilmark{1}}
\author{M. S. Brotherton, Wil van Breugel}
\affil{Institute of Geophysics and Planetary Physics, Lawrence Livermore National Laboratory, 7000 East Avenue, P.O. Box 808, L413, Livermore, CA 94550; mbrother@igpp.llnl.gov, wil@igpp.llnl.gov}

\author{R. J. Smith}
\affil{Mt. Stromlo \& Siding Spring Observatories, Australian National University, Private Bag, Weston Creek, ACT 2611, Australia; rsmith@mso.anu.edu.au}

\author{B. J. Boyle}
\affil{Anglo-Australian Telescope, PO Box 296, Epping, NSW 2121, Australia; director@aaoepp.aao.gov.au}

\author{T. Shanks, S. M. Croom}
\affil{Department of Physics, University of Durham, South Road, Durham, DH1 3LE, UK}

\author{Lance Miller}
\affil{Department of Physics, Oxford University, Keble Road, Oxford, OX1 3RH, UK; l.miller1@physics.ox.ac.uk}

\author{Robert H. Becker}
\affil{University of California at Davis, Davis, CA 95616; and Institute of Geophysics and Planetary Physics, Lawrence Livermore National Laboratory, 7000 East Avenue, P.O. Box 808, L413, Livermore, CA 94550; bob@igpp.llnl.gov}

\altaffiltext{1}{Based on observations at the W. M. Keck Observatory.}

\vskip -1cm
%\vfil\eject
\begin{abstract}

We report the discovery of five broad-absorption-line (BAL) QSOs
in a complete sample of 111 ultraviolet excess (UVX) QSO candidates
also detected in the NRAO VLA Sky Survey.  
All five BAL QSOs, which include two high-ionization
BAL QSOs and three low-ionization BAL QSOs, are formally radio-loud.
Of QSOs with $z > 0.4$, 3$\pm$2\% show low-ionization BALs, and of QSOs with
$z > 1.5$, all radio-loud, 9$\pm$5\% show BALs; these frequencies are
consistent with those of optical surveys.
While the first reported radio-loud BAL QSO, FIRST J155633.8+351758,
is likely to be heavily dust reddened and thus less radio-loud than
indicated by its observed radio-to-optical luminosity, these QSOs are UVX
selected and probably free of significant dust along the line of sight.  
We point out unusual features in two of the BAL QSOs and discuss
the significance finding these heretofore rare objects.

\end{abstract}
\keywords{quasars: absorption lines, quasars: emission lines,
quasars: general}
\vfil\eject
\section{Introduction}

The nature of the central engines of AGNs and the mechanisms by which they
produce their enormous luminosities and radio jets are not yet well understood.
The power of these jets differs by orders of magnitude in quasars that appear
similar at UV and optical wavelengths.  The distribution of
radio-to-optical power ranges over nearly six orders of magnitude and appears
bimodal (e.g., Kellermann et al. 1989; Stocke et al. 1992; Hooper et al. 1995).
Why are only a fraction of quasars ``radio-loud?''  Why is this fraction about
a tenth?

One previously strong result that seemed to be a clue to understanding this 
radio-loud/radio-quiet dichotomy was that broad-absorption-line (BAL)
QSOs appeared to be exclusively radio-quiet (e.g., Stocke et al. 1992).
This suggested that outflows in radio-quiet QSOs are not well collimated and
either formed or accelerated BAL material, while outflows in radio-loud
QSOs are confined to well collimated relativistic jets.

Becker et al. (1997a) reported the discovery of an exception to the
above result in the radio-loud BAL QSO FIRST J155633.8+351758, hereafter 
FIRST 1556+3517.  While the radio flux is high ($S_{1400} = 30.6$ mJy) for its 
redshift (z=1.48), making its radio luminosity consistent with those of 
radio-loud QSOs, the BAL QSO appears optically reddened and relativistically
beamed in the radio, making it unclear just how radio-loud the object truly is.
Near-IR photometry shows that FIRST 1556+3517, with $B-K=6.57$, is
among the reddest QSOs known (Hall et al. 1997).  ISO observations are also 
consistent with significant dust causing a line-of-sight reddening of $A_V = 
1.6$, which indicates $4M_{\sun}$ of dust and 800 $M_{\sun}$ of 
associated gas (Clavel 1998); correcting the optical magnitude for the 
reddening would bring FIRST 1556+3517 within approximately 1 sigma of the 
radio-loud/radio-quiet division.   Furthermore, Brotherton et al. (1997)
showed that much of the blue light is scattered, making the 
direct light from FIRST 1556+3517 even more red than apparent.  

While just how radio-loud FIRST 1556+3517 is remains an open issue,
Becker et al. (1997b) report the discovery of other radio-loud BAL QSOs
from the FIRST Bright Quasar Survey (FBQS; Gregg et al. 1995), which
selects stellar objects to a limiting magnitude of $E \sim 17.5$ with
$O-E < 2.0$ that appear in the FIRST, a deep radio survey that 
reaches 1 mJy at 20 cm (Becker et al. 1995). 
While none of these new radio-loud BAL QSOs are as red as FIRST 1556+3517,
the FBQS has a weak color selection.  Therefore the prospect of reddening in
FBQS radio-loud BAL QSOs remains an issue, especially because none are 
overly powerful radio sources. 

As a result of the discovery of these new objects, Weymann (1997) has 
refined the statement of the anticorrelation between radio-loudness and
the exhibition of BALs:
``The observed radial terminal velocity of thermal gas being ejected from
luminous QSOs is strongly anticorrelated with the radio power of the QSO.''
BAL QSOs that are formally radio-loud are interesting transition objects,
which may display related properties that will shed light on the
radio-loud/radio-quiet dichotomy. 

In this Letter, we report the discovery of five radio-loud BAL QSOs, selected
as stellar objects with an ultraviolet excess (UVX), and detected by the 
NRAO VLA Sky Survey (NVSS; Condon et al. 1998),
within a complete sample of 111 QSO candidates.  Because of the 
UVX selection, these QSOs are unlikely to be reddened significantly.
Section 2 describes the sample selection and observations.  Section 3 
tabulates the new BAL QSOs and their properties and gives their statistics
with respect to the rest of the sample.  Section 4 discusses the potential
importance of these BAL QSOs and how our results compare with other surveys.
As recommended by Weymann (1997), we adopt log R* as the formal measure
of radio-loudness with log R* = 1.0 dividing radio-loud and radio-quiet
QSOs, where R* is the K-corrected ratio of radio-to-optical power 
(Sramek \& Weedman 1980; Stocke et al. 1992).

\section{The Sample and Observations}

The UVX-NVSS sample includes quasar candidates from the UVX (AAT-2dF)
quasar survey (Smith et al. 1996) that are within 10$\arcsec$
of a radio source from the NVSS.
To make the matches, we extracted our own radio catalog from the NVSS images
using the HAPPY algorithm (White et al. 1997).

The UVX catalog includes just over 46000 stellar objects that meet the following
selection criteria.  The primary color selection is $U-B_j \leq -0.12$, 
which accounts for 90\% of the QSO candidates.  
Additional color selection using $R$\ extends the redshift range of
QSOs selected to $2.2 < z < 2.9$, about 10\% of the catalog.
The magnitude limits are 18.25 $< B_j <$ 20.85.
The area surveyed covers approximately 740 square degrees in
two strips: (1) between declination +2.5$\arcdeg$ and $-$2.5$\arcdeg$ and
right ascension 9.8 hours to 14.8 hours and (2) between declination 
$-$27.5$\arcdeg$ and $-$32.5$\arcdeg$ and right ascension 21.7 hours to 3.3 
hours.  Preliminary 2dF observations demonstrate that the UVX catalog is
approximately 50\% contaminated by non-AGN sources, predominantly stars and 
narrow-line galaxies.  The additional restriction that candidates also 
have a 20 cm radio flux density $S_{20cm} \gtrsim$ 2.5 mJy (5 $\sigma$ limit) 
reduces the contamination to $\sim5\%$ based on results below. 
The $z<2.2$ QSO sample is expected to be $>90\%$ complete with respect to 
previous QSO surveys, with a lower level of completeness for $2.2 < z < 2.9$;
additional details of the multicolor selection and completeness will
be given by Smith et al. (1998).

We obtained spectra of all 111 quasar candidates
from the equatorial UVX-NVSS catalog with $18.25 \leq B_j \leq 20.00$ and 
right ascensions from 10 hours to 13 hours.  Observations were
conducted with the Keck II telescope during 1998 February 9 and 
1998 March 7 (UT), using the Low Resolution Imaging Spectrometer
(Oke et al. 1995). The 300 line mm$^{-1}$ grating blazed at 5000 \AA\ with a 
1$\arcsec$ slit gave an effective resolution $\leq$ 10 \AA\ 
(FWHM of comparison lamp lines); the dispersion was 2.5 \AA\ pixel$^{-1}$ 
covering approximately 4000 \AA\ to 9000 \AA.  
All exposure times were four minutes.
%giving a signal-to-noise ratio per pixel of 10 to 50.
We employed standard data reduction techniques within the NOAO IRAF package.

\section{Results}

Of the 111 candidates, 103 are quasars (93\%), five are BL Lacertae objects, one
is a Seyfert 2 galaxy, and two are stars.  Given our observed wavelength
range, low-ionization BALs are visible when present in quasars with $z> 0.4$ 
(Mg II $\lambda2800$ enters the window; 101 objects), and high-ionization BALs
are visible when present in quasars with $z \gtrsim 1.5$ 
(C IV $\lambda$1549 enters the window; 33 objects).  
Given the selection criteria, all quasars in our sample with $z> 1.5$ will be
formally radio-loud (log R* $>$ 1); in the case of these specific 
objects, due to the combination of the properties of the quasar population, 
the selection criteria, and chance, all 103 quasars are formally radio-loud.  
%Uncertainties given below are standard deviations assuming a binomial 
%distribution.

We find three low-ionization BAL QSOs (3$\pm$2\% of the quasars with $z> 0.4$)
and two high-ionization BAL QSOs (6$\pm$4\% of the quasars with $z> 1.5$; one
of the low-ionization BAL QSOs also has $z> 1.5$ making a total of 
9$\pm$5\% BAL QSOs in the high-$z$ category).  
All are formally radio-loud, three solidly with
log R* $\gtrsim$ 2.  Figure 1 shows their spectra, and Table 1 lists
their properties, including the means and standard deviations for the
sample.  As seen in the radio-loud BAL QSOs found in the FBQS,
none are powerful radio sources.  Table 2 summarizes the BAL properties. 
All BAL identifications are
unambiguous, as the velocity widths ($V_{max} - V_{min}$) 
are $> 2000$\ km s$^{-1}$\ and, except 
for 1252+0053, we observe troughs of multiple species.  All the QSOs
have a positive ``balnicity,'' a conservative definition of a BAL QSO
(Weymann et al. 1991), although a number of lower velocity 
dispersion and associated systems are now recognized to be intrinsic in
a handful of QSOs.

\def\e{$\pm$}
\begin{deluxetable}{lcccccccccc}
\footnotesize
\tablewidth{0pt}
\tablenum{1}
\tablecaption{Radio-loud UVX BAL QSO Properties\tablenotemark{a}} 
\tablehead{Name & R.A. & Decl. & z & $B_j$\tablenotemark{b} & $U-B_j$\tablenotemark{b} & $B_j-R$\tablenotemark{b} & $M_B$ & $S_{1400}$\tablenotemark{c} & log $L_{14
00}$ & log R*\tablenotemark{d} \nl
 & (J2000) & (J2000) & & & & & & (mJy) & (ergs s$^{-1}$ Hz$^{-1}$) & }
\startdata
UN J1053$-$0058 & 10 53 52.83 & $-$00 58 53.63 & 1.55 & 18.72 & $-$0.20 & 0.87 & $-26.8$ & 23.4\e1.2 & 33.5 & 1.98 \nl
UN J1104$-$0004 & 11 04 40.81 & $-$00 04 42.55 & 1.35 & 19.55 & $-$0.32 & 0.76 & $-25.6$ & 33.1\e1.4 & 33.5 & 2.49 \nl
UN J1141$-$0141 & 11 41 06.54 & $-$01 41 08.48 & 1.27 & 19.84 & $-$0.83 & 0.39 & $-25.1$ &  \phn4.3\e0.5 & 32.6 & 1.73 \nl
UN J1225$-$0150 & 12 25 23.18 & $-$01 50 35.02 & 2.04 & 19.37 & $-$0.96 & 1.16 & $-26.8$ &  \phn3.6\e0.5 & 32.9 & 1.38 \nl
UN J1252$+$0053 & 12 52 43.78 & $+$00 53 19.69 & 1.69 & 19.22 & $-$0.76 & 0.49 & $-26.5$ & 16.3\e1.1 & 33.4 & 2.01 \nl
&  &  &  &  &  &  &  &  &  &  \nl
Sample\tablenotemark{e} Mean & 11.5 hr & \phs$0.0\arcdeg$ & 1.4 & 19.3 & $-$0.53 & \phs0.45 & $-25.8$ & \phn116 & 33.5 &  2.41 \nl
\hfil\hfil Std. Dev.& \nodata & \nodata & 0.6 & \phn0.4 & \phs0.42 & \phs0.36 & \phs\phn1.1 & \phn334 & \phn0.7 & 0.64 \nl
\hfil\hfil Minimum & 10\phd\phn\phn hr & $-2.5\arcdeg$ & 0.4 & 18.5 & $-$1.66 & $-$0.70 & $-$22.2 & \phn\phn\phn3 & 31.8 & 1.25 \nl
\hfil\hfil Maximum & 13\phd\phn\phn hr & $+2.5\arcdeg$ & 2.9 & 20.0 & \phs0.99 & \phs1.16 & $-$28.1 & 3092 & 35.7 & 4.13 \nl
\enddata
\tablenotetext{a}{Assuming $H_0 = 50, q_0 = 0, \alpha_{rad} = -0.1, \alpha_{opt}
 = 0.3, (S_{\nu} \propto \nu^{\alpha})$.}
\tablenotetext{b}{UVX Survey (Smith et al. 1996).  We have used $B=B_j-0.1$\ for
calculating $M_B$ and log R*.}
\tablenotetext{c}{NVSS (Condon et al. 1998).}
\tablenotetext{d}{For ease of comparison, as defined by Sramek \& Weedman 
(1980) and extensively used by Stocke et al. (1992) with their choice of 
$\alpha_{rad} = -0.3$ and $\alpha_{opt} = -1.0$ for K-corrections.}
\tablenotetext{e}{Properties of the complete sample of 103 quasars.  Note that
several properties, like the radio flux, are not normally distributed.}
\end{deluxetable}
\normalsize

\begin{deluxetable}{lccccc}
\tablewidth{0pt}
\tablenum{2}
\tablecaption{BAL Properties\tablenotemark{a}}
\tablehead{Name & Line & EW\tablenotemark{a} & Balnicity\tablenotemark{b} & $V_{min}$ & $V_{max}$ \nl
 & & (\AA) & (km s$^{-1}$) & (km s$^{-1}$) & (km s$^{-1}$) }
\startdata
UN J1053$-$0058 & C IV & 14 & 255 & 2100 & $-$5300 \nl
                & Al III & 5 & 50 & 1400 & $-$3600 \nl
                & Mg II & 5 & 40 & 1100 & $-$4600 \nl
UN J1104$-$0004 & Al III & 13 & 1100 & $-$800 & $-$6000 \nl
                & Mg II & 9 & 400 & $-$1100 & $-$5000 \nl
UN J1141$-$0141 & Al III & 4 & 400 & $-$1500 & $-$5200 \nl
                & Mg II & 1 & 170 & $-$1100 & $-$5000 \nl
UN J1225$-$0150 & C IV & 23 & 3900 & $-$8400 & $-$24300 \nl
                & Si IV & 12 & 2300 & $-$8000 & $-$22000 \nl
UN J1252$+$0053 & C IV & 5 & 130 & $-$8400 & $-$15400 \nl
\enddata
\tablenotetext{a}{Rest-frame lower limit based on a linear interpolation.}
\tablenotetext{b}{Defined by Weymann et al. (1991) who tabulate many
balnicities for C IV.}
\end{deluxetable}

\psfig{file=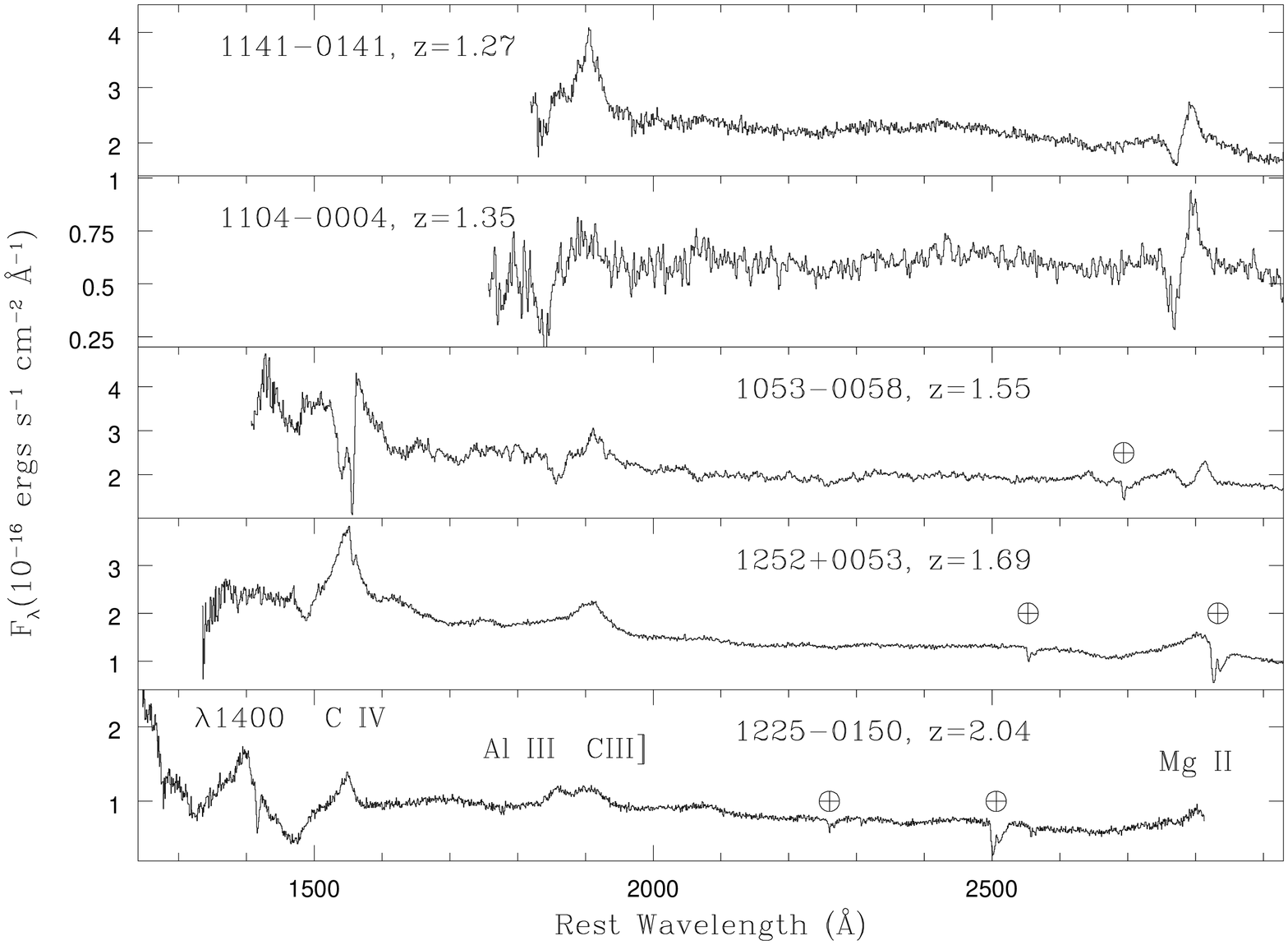,height=13cm,width=17cm,angle=-90}
\figcaption{Rest-frame spectra of the UVX radio-loud BAL QSOs.  Telluric A and
B band absorption has been marked ($\oplus$), and features of interest
are labeled in the bottom panel.  The spectrum of 1104$-$0004 has been smoothed
by 3 pixels.  Note that the bottoms of the flux scales do not start at zero.}

\section{Discussion}

\subsection{Individual Features}

There are interesting features in two BAL QSOs.  First,
the absorption profiles in 1053$-$0058 are unusual, appearing to extend to
the red of
the emission peak as well as to the customary blue: based on the redshift of 
the peak of \ion{C}{3}] $\lambda1909$, the \ion{Mg}{2} $\lambda2800$\ emission
should peak near the center of the absorption, and the \ion{C}{4} $\lambda1549$\
emission should peak near the local maximum in its corresponding absorption 
trough.  This suggests an unusual flow pattern or significantly blueshifted
emission lines (e.g., Brotherton et al. 1994).  

Second, in 1225$-$0150 the emission-line spectrum is unusual and there
is a ``mini-BAL'' candidate.  The intensity ratio of \ion{Al}{3} 
$\lambda1857$\ to \ion{C}{3}] $\lambda1909$ is 0.88 (based on a two-Gaussian 
fit), significantly larger than nearly all other known QSOs, including other 
BAL QSOs, which have enhanced \ion{Al}{3}/\ion{C}{3}] compared to radio-quiet
QSOs (Weymann et al. 1991).  (An exception might be the handful of QSOs like
H0335$-$336 (Hartig \& Baldwin 1986), which show strong \ion{Al}{3} 
$\lambda1857$\ and a strong \ion{Fe}{3} in place of \ion{C}{3}] $\lambda1909$.)
\ion{C}{4} $\lambda$1549 emission appears weak relative to emission from
\ion{C}{3}] $\lambda1909$ and the $\lambda$1400 feature 
(a blend of \ion{Si}{4} $\lambda$1397 and \ion{O}{4}] $\lambda$1402; 
see Wills et al. 1993); the \ion{C}{4} $\lambda$1549 emission line does not 
appear to be affected by the BAL trough, which appears detached by more than
8000 km s$^{-1}$.  
What these unusual line ratios mean is not 
clear, although strong \ion{Al}{3} $\lambda1857$\ is an important coolant
for very dense gas (e.g., Ferland et al. 1996). Because \ion{C}{4} 
$\lambda$1549 is likely to have a very high optical depth, 
the line may become an
ineffective emitter at high density (e.g., Zheng \& Puetter 1990).
Additionally, there is a smooth, intermediate width ($\sim$ 900 km s$^{-1}$) 
absorption system at $-$27,000 km s$^{-1}$ seen in both \ion{C}{4} 
$\lambda$1549 and \ion{Si}{4} $\lambda$1397 (the latter line is not 
commonly seen in interstellar systems); we suggest that this may be an 
intrinsic ``mini-BAL'' system (e.g., Hamann et al. 1997), but higher
resolution spectroscopy, perhaps at multiple epochs, would be needed to
confirm our suggestion.

\subsection{Sample Features}

%In the following discussion, the uncertainty associated with small number 
%statistics should be kept in mind.  
Foltz et al. (1990) find that 9\% of the
optically selected Large Bright Quasar Survey (LBQS;
Foltz et al. 1987, 1989; Hewett et al. 1991; Chaffee et al. 1991; Morris et al.
1991; Francis et al. 1992) QSOs with $z>1.5$ are BAL QSOs, the same fraction
we find for our $z>1.5$ radio-loud UVX sample.  One of the three
BAL QSOs in our high-redshift subsample displays low-ionization BALs.
While this result is not statistically unlikely given the expected 0.1 
fraction of low-ionization BALs in optically selected samples, it is 
suggestive of an excess because this is a strongly color-selected sample.
Previous work has shown that low-ionization BAL QSOs tend to be more dust 
reddened than either high-ionization BAL QSOs or the bulk of radio-quiet QSOs
(Sprayberry \& Foltz 1992).

The question remains why radio-loud BAL QSOs were not identified sooner.
Two methods have been employed: searching the spectra of radio-loud QSOs 
for BALs, and measuring the radio-loudness of optically selected BAL QSOs
(e.g., Stocke et al. 1992).  The first method failed apparently because
of the shallowness of radio surveys given the persistent but unexplained
anticorrelation stated in section 1.  While our results suggest that
the incidence of BALs is independent of whether log R* is greater than
or less than unity, it appears that there is some value of log R* above
which BALs are not seen.  The answer to this issue will likely emerge
with the growth of the FBQS (e.g., Gregg et al. 1995).
The best current limits are discussed by Weymann (1997), who finds no BALs
among 20 QSOs with 6 cm flux densities between 30 and 100 mJy.  The radio-loud
BAL QSOs reported here and by Becker et al. (1997b) have 20 cm flux densities
of no higher than approximately 30 mJy.  Given the optical fluxes of the QSOs,
the radio-loudness above which BALs are not seen appears to be near to our 
largest log R* = 2.5.

The second method of discovering radio-loud QSOs, while searching
deep enough in the radio, failed apparently through insufficient numbers.
There are about 5200 UVX quasar candidates that meet the present selection 
criteria independent of NVSS detection, half of which should be QSOs, and 
five of which we found to be radio loud BAL QSOs.  Therefore, the fraction of
radio-loud BAL QSOs found with pure optical selection is at least 1 in 500
(at least because the NVSS does not detect all the radio-loud QSOs
in the sample, and because we do not know how many of the low-redshift QSOs
may show BALs in the unobserved ultraviolet).
Assuming that 1/3 of the $\sim$2600 UVX QSOs have $z>1.6$, for which the 
high-ionization \ion{C}{4} $\lambda$1549 BAL trough can be
seen at wavelengths greater than 4000 \AA\, the observed radio-loud BAL QSO 
fraction is greater than about 1 in 200.  At the lower limit of these 
detection rates, the LBQS was too small to be sure of finding such an object. 
Because Becker et al. (1997b) find radio-loud BAL QSOs within the FBQS
(Gregg et al. 1995), the fact that the UVX-NVSS sample is optically
fainter than the LBQS probably is not important.

Despite differences in selection criterion, there are similarities that
bias both the FBQS and the UVX-NVSS sample toward finding radio-loud BAL QSOs.
Both samples preferentially select large numbers of radio-intermediate
and radio-loud but low log R* QSOs.  Furthermore, the radio selection is
at a relatively high frequency, 1400 MHz.  If Falcke et al. (1996) are
correct that radio-quiet QSOs possess relativistic radio-emitting jets
and that some fraction of radio-loud but low log R* QSOs are actually
beamed radio-quiet QSOs, the pure anticorrelation between radio-loudness
and the presence of BALs may still be preserved.  Radio maps with high
spatial resolution or spectral indices are needed for these radio-loud
BAL QSOs to rule out strong beaming.  Such a face-on geometry with 
BAL outflow along a radio-quiet jet axis could explain the overabundance
of BALs seen among radio-intermediate QSOs by Francis, Hooper, \& Impey (1993).

\section{Conclusions}

We have discovered five radio-loud BAL QSOs in a complete sample of 103
deep-radio detected UVX QSOs which are unlikely to be significantly dust
reddened.  The 9$\pm$5\% BAL QSOs in the high-redshift sample 
($z \gtrsim 1.5$, entirely radio-loud) is consistent with the frequency in 
purely optically selected QSOs (9\%).
We find three (3$\pm$2\%) low-ionization BAL QSOs, also consistent with 
pure optical selection ($\sim1\%$).  
While BALs are not seen in the most radio-loud QSOs, the incidence of BALs 
appears to be independent across the conventional radio-loud/radio-quiet 
division.

\acknowledgments

We thank Carlos De Breuck and the Keck staff
%Ron Quick, Randy Campbell, Tom Bida, Terry Stickel, and Bob Goodrich
for their excellent assistance.  We thank Nahum Arav and Fred Hamann
for their comments on the manuscript.
The W. M. Keck Observatory is a scientific partnership between the
University of California and the California Institute of Technology,
made possible by the generous gift of the W. M. Keck Foundation.
This work has been performed under the auspices of the U.S. Department of Energy
by Lawrence Livermore National Laboratory under Contract W-7405-ENG-48.

%\clearpage
%tables go in here separated by \clearpage statements.

\end{document}